\documentclass[aps,preprint, nofootinbib]{revtex4}
\usepackage{graphicx}
\usepackage{amssymb}

\usepackage{color}

\def\bea{\begin{eqnarray}}
\def\eea{\end{eqnarray}}

\makeatletter

\begin{document}

\begin{flushright}
ANL-HEP-PR-10-51\\ 
NUHEP-TH/10-21\\
UCI-HEP-TR-2010-20\\
FERMILAB-PUB-10-371-T\\
\end{flushright}

\title{Gamma Ray Lines from a Universal Extra Dimension}
\author{Gianfranco Bertone$^a$, C.~B. Jackson$^{b}$, \\
Gabe Shaughnessy$^{c,d}$, Tim M.P. Tait$^{e}$ and Alberto Vallinotto$^{f}$}
\affiliation{
\mbox{$^a$ Institut for Theoretical Physics, Univ. of Z\"urich, Winterthurerst. 190, 8057 Z\"urich CH,} \\
\mbox{IAP, UMR 7095-CNRS, Univ. P. et M. Curie, 98bis Bd Arago, 75014 Paris, France}
\mbox{$^b$Department of Physics, University of Texas at Arlington, Arlington, TX 76019} \\
\mbox{$^c$High Energy Physics Division, Argonne National Laboratory, Argonne, IL 60439}  \\ 
\mbox{$^d$Northwestern University, 2145 Sheridan Road, Evanston, IL 60208}\\
\mbox{$^e$Department of Physics and Astronomy, University of California, Irvine, CA 92697}\\
\mbox{$^f$Center for Particle Astrophysics, Fermi National Accelerator Laboratory, Batavia, IL 60510}
}

\vspace*{-0.5cm}

\begin{abstract}
Indirect Dark Matter searches are based on the observation of secondary particles produced by the annihilation or decay of Dark Matter.
Among them, gamma-rays are perhaps the most promising messengers, as they do not suffer deflection or absorption on Galactic 
scales, so their observation would directly reveal the position and the energy spectrum of the emitting source. 
Here, we study the detailed gamma-ray energy spectrum of Kaluza--Klein Dark Matter in
a theory with 5 Universal Extra Dimensions. We focus in particular on the two body annihilation of Dark Matter particles into a 
photon and another particle, which produces monochromatic
photons, resulting in a line in the energy spectrum of gamma rays. Previous calculations in the context of the five dimensional UED model have computed the
line signal from annihilations into $\gamma \gamma$, but we extend
these results to include $\gamma Z$ and $\gamma H$ final states.  We find that these spectral lines are subdominant compared
to the predicted $\gamma \gamma$ signal, but they would be important as follow-up signals
in the event of the observation of the $\gamma \gamma$ line, in order to distinguish
the 5d UED model from other theoretical scenarios. 
\end{abstract}


\maketitle

\section{introduction}
\label{sec:intro}

There is overwhelming evidence that our Universe contains a large component of 
non-baryonic dark matter. However, so far its identity and nature have remained 
elusive \cite{Bertone:2004pz}.  Understanding
dark matter in a larger context remains one of the most compelling mysteries in particle
astrophysics.  With the advent of the current generation of gamma ray and neutrino observatories,
the next generation of direct detection experiments, and the successful operation of the Large Hadron Collider, we have
entered a promising era for the detection of dark matter through non-gravitational
interactions with the Standard Model (SM).

One of the deep mysteries of dark matter is the fact that on the one hand it must be massive,
but on the other, incredibly stable, with a lifetime on the order of the age of the Universe itself.
This odd combination of features argues for the presence of a symmetry which 
(at least to very good approximation) forbids the dark matter from decaying.  
The most straight-forward realization is a symmetry requiring
dark matter to couple in pairs with the Standard Model and we can classify theories of
dark matter based on how they realize this symmetry.

The most well known example of such a symmetry is the $R$-parity often built into
supersymmetric extensions of the SM to forbid dangerous
interactions leading to large baryon- and lepton-number violation.  While the fact that
$R$-parity leads naturally to a weakly interacting massive particle (WIMP) that can play
the role of dark matter makes it a very attractive ingredient, the existence of
$R$-parity itself is somewhat ad hoc.  It would be preferable to have a symmetry whose origin
has a deeper motivation.

A wide class of theories with extra dimensions naturally have such a symmetry.  The existence of
extra dimensions leads to additional spacetime symmetries, some remnant of which may survive
compactification of the extra dimensions.  In Universal Extra Dimensions (UED), the entire
Standard Model lives equally in all dimensions.  As a result, there is naively a translational
invariance along the extra dimensional directions which translates into conservation of
Kaluza Klein (KK) number in the interactions of the modes \cite{Appelquist:2000nn}.  
In practice, the need for chiral fermions in the low energy theory
selects orbifold compactifications which break the conservation of KK number into a $Z_2$
symmetry, but this is sufficient to guarantee a stable particle which can play the role
of dark matter.

The simplest example of UED is a model with five dimensions (5d), 
four of which are ordinary spacetime with the remaining one compactified
into a line segment of length $L$.  The lightest Kaluza Klein (KK) mode
is stable, and provided it is weakly interacting, can take the role of dark matter in the
Universe.  A standard thermal history of the Universe results in the correct relic density
for masses on the order of several hundred GeV \cite{Servant:2002aq,Kong:2005hn},
a region which is also consistent with Tevatron and precision
electroweak bounds \cite{Appelquist:2002wb,Cheng:2002ab} 
and a discovery of at least part of the
first level of KK modes at the LHC \cite{Cheng:2002ab,Datta:2005zs}.  There are
also prospects for direct \cite{Cheng:2002ej,Servant:2002hb} and indirect
\cite{Baltz:2004ie,Hooper:2002gs,Barrau:2005au,Bertone:2002ms,Bergstrom:2004nr}
detection by experiments in the near future.

In this article, we focus in particular on the prospects for detecting gamma rays from annihilation
of WIMPs in the galactic halo.  Gamma rays offer a particularly promising avenue to detect
WIMP annihilations, because they point back to their sources, unlike charged particles, which
are at the mercy of the galactic magnetic fields.  However, the backgrounds from gamma
ray sources are not very well understood, particularly around the galactic center where the
concentration of WIMPs is greatest.  Thus, we focus on the particular signal of a two body
annihilation of two WIMPs into a photon and another particle, which produces monochromatic
photons, resulting in a ``line" in the energy spectrum of gamma rays.  This feature
is sufficiently distinct from anything produced by conventional astrophysics so as to perhaps
balance the smaller rate caused by the fact that it is a loop-induced process \cite{Mack:2008wu}.
The Fermi gamma ray observatory is actively searching for such lines
as an unequivocal signal of dark matter \cite{Abdo:2010nc}, and predictions from
a wide variety of particle physics models have appeared in the literature
\cite{Bertone:2009cb,Bergstrom:1997fh,Dudas:2009uq,Perelstein:2006bq,Goodman:2010qn}.

Previous calculations in the context of the five dimensional UED model have computed the
line signal from WIMPs annihilating into $\gamma \gamma$ \cite{Bergstrom:2004nr}.  We extend
these results to include $\gamma Z$ and $\gamma H$ final states (and we independently
verify the results of \cite{Bergstrom:2004nr}, finding agreement for the $\gamma \gamma$
channel).  While both $\gamma Z$ and $\gamma H$ turn out to be subdominant compared
to the predicted $\gamma \gamma$ signal, they would be important as follow-up signals
in the event of the observation of the $\gamma \gamma$ line, in order to distinguish
the 5d UED model from other possibilities.  For example, the relative strengths of the
$\gamma \gamma$ and $\gamma Z$ lines encodes the fact that the LKP couples proportionally
to hypercharge, and the existence of a $\gamma H$ line, in addition to being interesting in its own
right, is only expected to be visible for a WIMP which is a vector (such as the LKP) or
a Dirac fermion.

We proceed as follows.  In Section~\ref{sec:model}, we briefly review the 5 dimensional UED
model, and the couplings necessary for our later computations.  Sections~\ref{sec:continuum}
and \ref{sec:lines} respectively cover the expected gamma ray continuum and lines resulting
from LKP annihilations.  Astrophysical inputs, the effects of finite detector energy resolution,
and the gamma ray spectra are assembled in Section~\ref{sec:flux}.  We conclude in
Section~\ref{sec:conclusions}.

\section{5d UED Model}
\label{sec:model}
 
 The 5d UED model \cite{Appelquist:2000nn}
 consists of the ordinary four large space + time dimensions, with one
 additional spatial dimension compactified into a line segment of length $L$.
 Generic points in five dimensional space can be written $x^M \equiv (x^\mu, y)$, where
 $x^\mu$ is the subspace of the four large spacetime dimensions and $0 \leq y \leq L$ is
 the compact dimension.
 The Standard Model field content is promoted to functions of the complete spacetime $x^M$,
 with the zero modes in the Kaluza-Klein expansion identified with the familiar SM fields.  In
 the discussion below, we will interchangeably refer to the ordinary SM fields as either
 ``SM fields" or ``zero-modes".  After electroweak symmetry-breaking, the lightest states
 obtain the usual SM masses, but we nonetheless
 continue to refer to them as zero modes in the Kaluza-Klein sense.
 
Orbifold  boundary conditions project out the zero modes of the fifth components of the gauge
fields (the higher components are eaten by the KK gauge bosons as per 
the usual extra-dimensional Higgs mechanism) and the unwanted zero mode degrees
of freedom of the fermions.  The boundary conditions lead inevitably
\cite{Georgi:2000ks} to the presence of brane-localized terms which
further shift the masses of the higher KK modes and break translational invariance along the
extra dimension into a discrete parity under which the odd-numbered KK modes are odd
and the even-numbered modes are even \cite{Cheng:2002iz,Carena:2002me}.  As a consequence
of this $Z_2$ remnant symmetry, the lightest first level Kaluza-Klein particle (LKP) is stable.

We follow
the general picture of ``minimal UED"  (mUED)
\cite{Cheng:2002ab,Cheng:2002iz} in which the brane terms take
sizes which are dictated by loops of bulk couplings, but we do not strictly restrict ourselves
to the masses and couplings of mUED\footnote{For a few interesting examples of more severe
deviations from mUED, see Refs.~\cite{Flacke:2008ne}.}.  
Consequently, the mass spectrum is chiefly
characterized by the KK number $n$, with the masses of all of the particles at that level
given roughly by
\bea
M_n \simeq \frac{\pi n}{L} .
\eea
The boundary terms lead to corrections for the various particles within a given level.  mUED
predicts the colored particles receive the largest positive corrections to their masses, followed
by the $SU(2)$-charged states.  The lightest KK modes of a given level are typically 
the KK modes of the right-handed charged leptons and the $U(1)$ gauge boson itself.
The LKP is thus the first level KK mode of $B_{\mu}$, and represents a vector dark matter
particle (often refered to as
the ``KK photon") which is a SM gauge singlet.  Its principle interactions are to
a first level KK fermion and its corresponding zero mode, proportional to the
hypercharge of the fermion in question and the gauge coupling $g_Y$.

The primary couplings of interest here are the couplings of the LKP to one KK fermion
and one zero mode fermion, and the couplings of ordinary photons, $Z$ bosons, and Higgs
bosons to either zero mode or KK fermions.  The photon couples to a pair of either
SM or KK fermions proportional to $e Q$ where $Q$ is the electric charge as usual.  
The SM Higgs couples to a pair of zero modes proportionally to their mass, or to
a pair of KK fermions (one of which must be an $SU(2)$ doublet and the other an
$SU(2)$ singlet) proportionally to the mass of the analogue zero mode fermion.

The SM $Z$ boson has the standard chiral couplings to the SM fermions, and vector-like
couplings to singlet ($s$) and doublet ($d$) KK fermions given by:
\bea
-\frac{e}{4 s_w c_w} \left(g_V + g_A \right) Z_\mu \bar{\xi_s}^{(1)} \gamma^\mu \xi_s^{(1)}
\label{eq:Zxixi_s}
\eea
and:
\bea
-\frac{e}{4 s_w c_w} \left(g_V - g_A \right) Z_\mu \bar{\xi_d}^{(1)} \gamma^\mu \xi_d^{(1)}
\label{eq:Zxixi_d}
\eea
where $\xi_d^{(1)}$ is the first level KK mode of a SM $SU(2)$ doublet, and $\xi_s^{(1)}$ is the
first level KK mode of a SM singlet.  $s_w (c_w)$ are the sine (cosine) of the 
weak mixing angle and $g_V$ and $g_A$ are defined as:
\begin{eqnarray}
g_V &=& -2 T_3 + 4 Q_f s_w^2 \\
g_A &=& 2 T_3\,.
\end{eqnarray}
The $B^{(1)}$ couples to one SM fermion and one KK fermion as:
\begin{equation}
-g_Y \frac{Y_s}{2} B_\mu^{(1)}  \bar{\xi_s}^{(1)} \gamma^\mu \left( 1 + \gamma_5 \right) \psi^{(0)} + c.c.
\label{eq:B1-xis-psi}
\end{equation}
and:
\begin{equation}
-g_Y \frac{Y_d}{2} B_\mu^{(1)}  \bar{\xi_d}^{(1)} \gamma^\mu \left( 1 - \gamma_5 \right) \psi^{(0)} + c.c.
\label{eq:B1-xid-psi} \,.
\end{equation}
where $\psi^{(0)}$ is the zero mode fermion.

\section{Continuum Gamma Rays}
\label{sec:continuum}

\begin{figure*}[t]
\begin{center}
\includegraphics[scale=0.8]{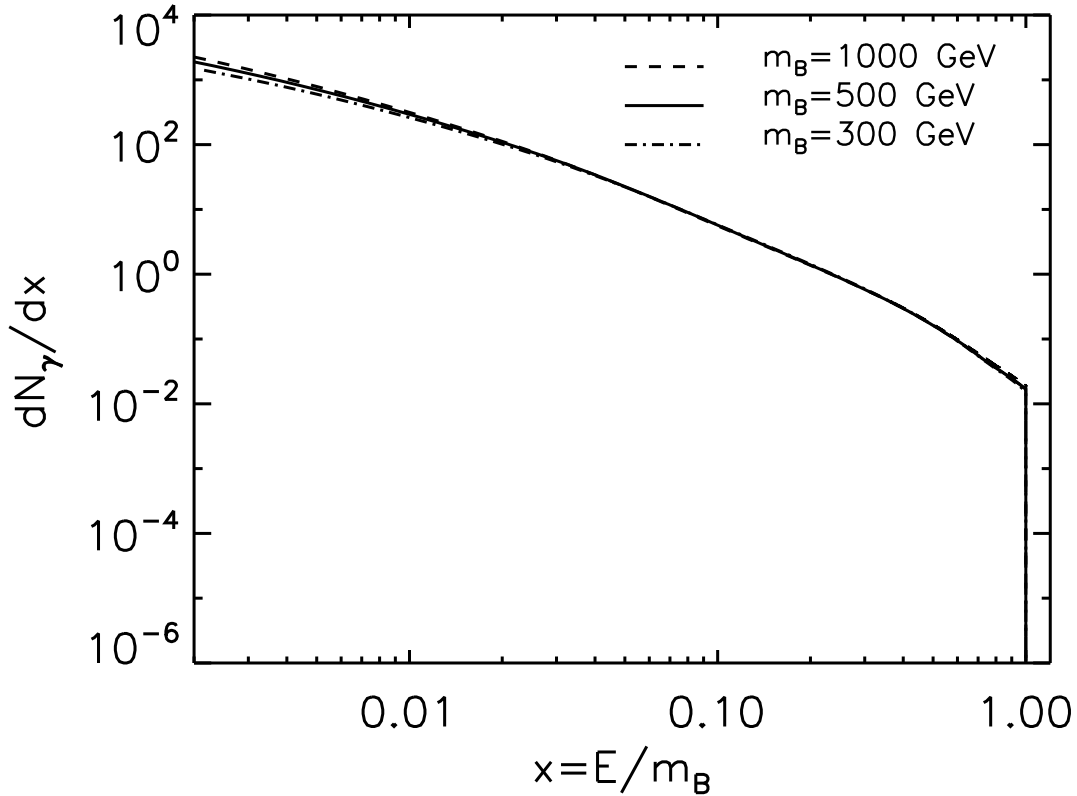}
\end{center}
\caption[]{The continuum gamma ray spectrum as a function of $x$, the ratio of the photon 
energy to the LKP mass, for LKP masses $m_B = 300$~GeV (dash-dotted curve), 
$500$~GeV (dashed curve), and 1 TeV (solid curve).}
\label{fg:continuum}
\end{figure*} 

Annihilations of two LKPs is predominantly into charged leptons ($\sim 59\%$),
with significant fractions into quarks ($\sim 35\%$), neutrinos ($\sim 4\%$), and Higgs bosons
($\sim 2\%$) \cite{Servant:2002aq}.
The fact that a large fraction of annihilations produces hard charged leptons has the
consequence that the continuum gamma ray signal is itself rather 
hard \cite{Cheng:2002ej,Bertone:2002ms}, which distinguishes
the 5d continuum spectrum from the six dimensional chiral square 
model \cite{Bertone:2009cb}.  The resulting continuum spectrum is shown in
Fig.~\ref{fg:continuum} as a function of $x \equiv E_\gamma / M_B$, the fraction of
energy of the final state photon normalized to the LKP mass.  Comparing curves for
LKP masses of 300, 500, and 1000~TeV, we see little difference in the photon spectrum,
which is easily explained by the fact that the annihilations are almost entirely into
light SM particles.

\section{Gamma ray line cross sections}
\label{sec:lines}

In addition to a diffuse continuum of gamma rays, WIMP annihilations are also expected to produce prompt photons via loop-level processes.  These types of annihilations produce $\gamma + X$ final states (where $X$ can be either a vector gauge boson or a scalar) and result in mono-energetic ``lines'' superimposed on the continuum.  The energy of these lines are determined almost solely by the mass of the WIMP and the $X$ particle:
\begin{equation}
E_\gamma =  M_{WIMP} \left( 1 - \frac{M_X^2}{4 M_{WIMP}^2} \right) \,.
\end{equation}
Due to the non-relativistic nature of WIMPs, the possible final states (i.e., the identity of $X$) are determined by the spin of the DM particle.  In the case of the 5d UED model, the WIMP is a vector gauge boson and, hence, $X$ can be either a vector or a scalar.  In other words, WIMP annihilations in the 5d UED model are capable of producing $\gamma\gamma$, $\gamma Z$ and (if kinematically-accessible) $\gamma H$ final states (where $H$ is the SM Higgs boson).  

The production of the $\gamma\gamma$ final state in the 5d UED model was first considered in Ref~\cite{Bergstrom:2004nr}.  Here, we will focus on the other two possible final states ($\gamma Z$ and $\gamma H$) and we refer the interested reader to the above reference for details on the calculation of the $\gamma\gamma$ cross section.  We have verified the results from the previous analysis and we present the numerical results for the flux in the following sections.

\begin{figure*}[t]
\begin{center}
\includegraphics[scale=0.6]{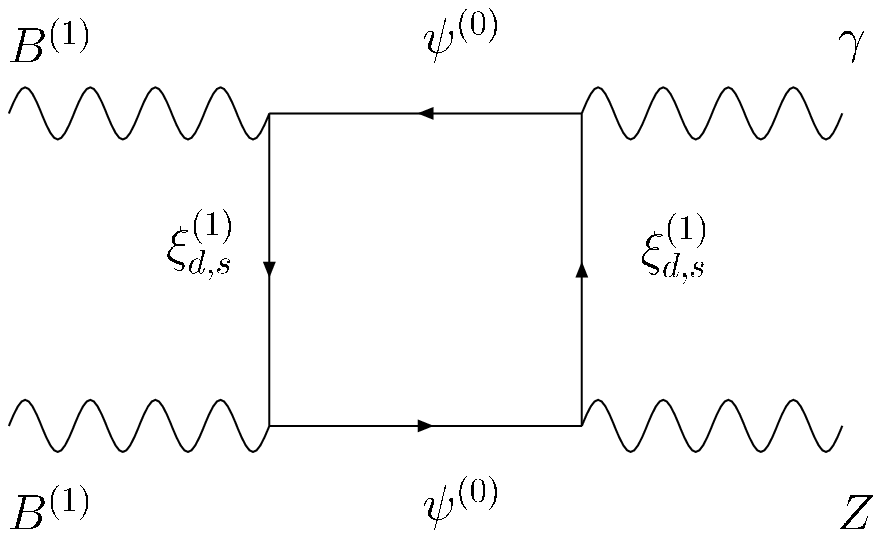} 
\includegraphics[scale=0.6]{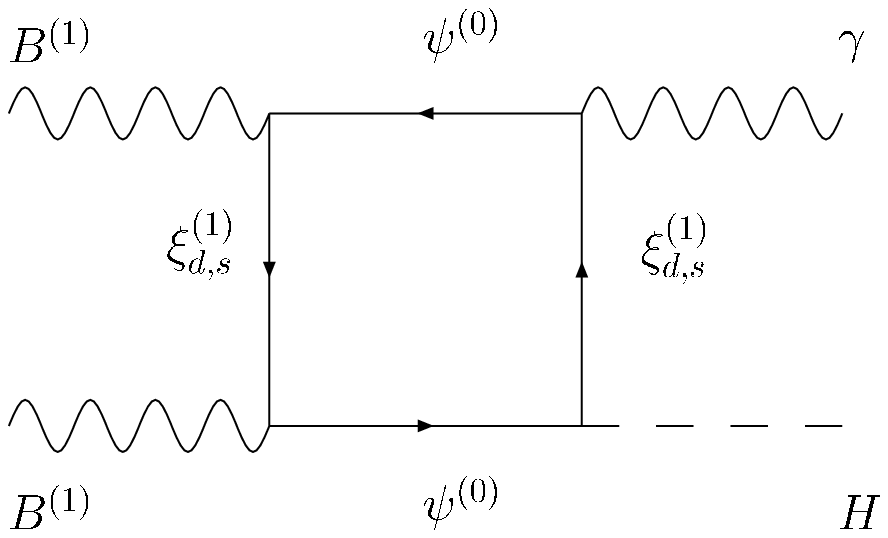} 
\end{center}
\caption[]{Examples of Feynman diagrams which contribute to (a) $B^{(1)} B^{(1)} \to \gamma + Z$ and (b) $B^{(1)} B^{(1)} \to \gamma + H$.}
\label{fg:feynman-diagrams}
\end{figure*}

Representative Feynman diagrams depicting the process $B^{(1)} + B^{(1)} \to \gamma + X$ 
(where $X = Z$ or $H$) are shown in Fig.~\ref{fg:feynman-diagrams}.  Note that since 
the LKP is to very good approximation pure $B^{(1)}$ ($\lesssim 10^{-1}$ for mUED
boundary terms and $L^{-1} \gtrsim 300$~GeV \cite{Cheng:2002iz}),
loops containing $W$ and/or KK charged Goldstone bosons are suppressed by the
tiny $W_3^{(1)}$ content.  We neglect these tiny contributions compared to the 
dominant diagrams consisting of closed loops of SM and KK fermions.

\subsection{Annihilation to $\gamma + Z$} 

For the $\gamma + Z$ process, we assign Lorentz indices and momenta as 
\bea
B^{(1)}_{\mu_1}(p_1) + B^{(1)}_{\mu_2}(p_2) \to \gamma_{\mu_3}(p_3) + Z_{\mu_4}(p_4).
\eea  
Following Ref.~\cite{Bergstrom:2004nr}, the scattering amplitude involving four external gauge 
bosons can be written as:
\begin{equation}
{\cal{M}} = \epsilon_1^{\mu_1}(p_1) \epsilon_2^{\mu_2}(p_2) \epsilon_3^{\mu_3 *}(p_3) \epsilon_4^{\mu_4 *}(p_4) {\cal{M}}_{\mu_1 \mu_2 \mu_3 \mu_4} (p_1, p_2, p_3, p_4)\,.
\end{equation}
In general, the subamplitude ${\cal{M}}_{\mu_1 \mu_2 \mu_3 \mu_4}$ can be expanded in terms of 
metric tensors and external momenta.  Taking into account the transversality of the polarization 
tensors ($\epsilon(p) \cdot p = 0$) and the non-relativistic nature of WIMPs today 
(such that $p_1 \simeq p_2 \equiv p$), the most general form is given by:
\begin{eqnarray}
{\cal{M}}^{\mu_1 \mu_2 \mu_3 \mu_4} &=& \frac{\alpha_Y \alpha_{em}}{4 c_w s_w} \sum_{\ell} Q_\ell \biggl[
\frac{A_\ell}{m_{B^{(1)}}^4} p_3^{\mu_1} p_4^{\mu_2} p^{\mu_3} p^{\mu_4} + \frac{B_{\ell, 1}}{m_{B^{(1)}}^2} g^{\mu_1 \mu_2} p^{\mu_3} p^{\mu_4} +  \frac{B_{\ell,2}}{m_{B^{(1)}}^2} g^{\mu_1 \mu_3} p_4^{\mu_2} p^{\mu_4} \nonumber\\
&&  \frac{B_{\ell,3}}{m_{B^{(1)}}^2} g^{\mu_1 \mu_4} p_4^{\mu_2} p^{\mu_3} +  \frac{B_{\ell,4}}{m_{B^{(1)}}^2} g^{\mu_2 \mu_3} p_3^{\mu_1} p^{\mu_4} +  \frac{B_{\ell,5}}{m_{B^{(1)}}^2} g^{\mu_2 \mu_4} p_3^{\mu_1} p^{\mu_3} \nonumber\\
&& +  \frac{B_{\ell, 6}}{m_{B^{(1)}}^2} g^{\mu_3 \mu_4} p_3^{\mu_1} p_4^{\mu_2}   + 
C_{\ell,1} g^{\mu_1 \mu_2} g^{\mu_3 \mu_4} + C_{\ell,2} g^{\mu_1 \mu_3} g^{\mu_2 \mu_4} + C_{\ell,3} g^{\mu_1 \mu_4} g^{\mu_2 \mu_3} \biggr] \nonumber\\
&& \equiv \frac{\alpha_Y \alpha_{em}}{4 c_w s_w} \sum_{\ell} Q_\ell {\cal{A}}_{ZA,\ell}^{\mu_1 \mu_2 \mu_3 \mu_4}\,.
\end{eqnarray}
where we have pulled out a common factor with $\alpha_Y = \alpha_{em}/c_w^2$ and we sum 
over all charged fermions running in the loop.
The fact that the WIMPs are annihilating nearly at rest 
(i.e., $p^\mu \simeq (m_{B^{(1)}}, \bf{0}))$ also 
allows one to define the $z$-axis in the center-of-mass frame as the axis along which the final state 
particles are traveling.  This simplifies things greatly as many of the dot products between 
polarization tensors and momentum vectors vanish.  Accounting for this and imposing conservation 
of momentum ($ 2 p = p_3 + p_4$), we find that the only remaining tensor structures are $B_2, B_4, 
B_6, C_1, C_2$ and $C_3$.

Computation of the loop integrals in WIMP annihilations is complicated by the fact that two of the incoming momenta are nearly identical.  In these configurations, usual approaches such as the Passarino-Veltman scheme \cite{Passarino:1978jh} for computing tensor integrals break down and one must rely on alternative schemes in order to safely compute one-loop scattering amplitudes.  We have chosen to use the scheme introduced in Ref.~\cite{Stuart:1987tt}.  This scheme was most recently used to compute one-loop WIMP annihilation scattering amplitudes and we refer the interested reader to in Ref.~\cite{Bertone:2009cb} for details.

Squaring the amplitude and summing/averaging over all polarizations, we find that the cross 
section for $B^{(1)} + B^{(1)} \to \gamma + Z$ takes the form:
\begin{equation}
\langle
\sigma_{\gamma Z} v \rangle 
= \frac{1}{9} \frac{\alpha_Y^2 \alpha_{em}^2}{16\pi c_w^2 s_w^2} 
\left( \frac{\beta_Z^2}{32 m_{B^{(1)}}^2} \right) \left| \sum_\ell Q_\ell {\cal{A}}_{ZA,\ell} \right|^2 \,,
\end{equation}
where $\beta_Z = \sqrt{1 - \frac{m_Z^2}{4 m_{B^{(1)}}^2}}$.

\subsection{Annihilation to $\gamma + H$}

The amplitude for this process takes the form:
\begin{equation}
{\cal{M}} = \epsilon_1^{\mu_1}(p_1) \epsilon_2^{\mu_2}(p_2) \epsilon_3^{\mu_3 *}(p_3)  {\cal{M}}_{\mu_1 \mu_2 \mu_3 } (p_1, p_2, p_3, p_4)\,
\end{equation}
where we have assigned the Lorentz indices and momenta as
\bea
B^{(1)}_{\mu_1}(p_1) + B^{(1)}_{\mu_2}(p_2) \to \gamma_{\mu_3}(p_3) + H(p_4).
\eea  
The amplitudes for this process involve one less vector boson, and thus have a simpler tensor
structure than the $\gamma Z$ case.   In fact,
due to the scalar coupling of the Higgs boson, the only surviving terms from the 
trace over the internal fermion line are proportional to Levi-Civita tensors $\epsilon^{\mu
\nu\lambda\sigma}$ contracted with at least one of the external momenta.  
Instead of expanding the scattering amplitude in terms of all possible 
permutations, we only list the surviving terms once the non-relativistic nature of the WIMPs and 
conservation of momentum are applied.  The resulting expression takes the form:
\begin{eqnarray}
{\cal{M}}_{\mu_1 \mu_2 \mu_3 } (p_1, p_2, p_3, p_4) &=&  \frac{\alpha_Y \alpha_{em}}{2 m_W s_w} \sum_{\ell} Q_\ell m_\ell \left(Y_{\ell, d}^2 - Y_{\ell, s}^2\right)  \biggl[ \frac{D_1}{m_{B^{(1)}}^3} p_A^{\mu_1} p_\sigma p_{A, \lambda} \epsilon^{\sigma \lambda \mu_2 \mu_3} \nonumber\\
&& + \frac{D_2}{m_{B^{(1)}}} p_\sigma \epsilon^{\sigma \mu_1 \mu_2 \mu_3} + \frac{D_3}{m_{B^{(1)}}} p_{A,\sigma} \epsilon^{\sigma \mu_1 \mu_2 \mu_3} \biggr] \nonumber\\
&& \equiv \frac{\alpha_Y \alpha_{em}}{2 m_W s_w} \sum_{\ell} Q_\ell m_\ell \left(Y_{\ell, d}^2 - Y_{\ell, s}^2\right) {\cal{A}}_{HA, \ell}^{\mu_1 \mu_2 \mu_3}\,.
\end{eqnarray}
We note that the scattering amplitude is directly proportional to the zero mode
fermion mass $m_\ell$ (via the Yukawa coupling).  As a result, only loops containing top quarks 
and their KK partners will make a significant contribution to the process.  Summing and averaging 
over all polarizations, we find the cross section can be written as:
\begin{equation}
\langle \sigma_{\gamma H} v \rangle 
= \frac{1}{9} \frac{\alpha_Y^2 \alpha_{em}^2}{4 \pi m_W^2 s_w^2} \left( \frac{\beta_H^2}{32 m_{B^{(1)}}^2} \right)  \left| \sum_\ell Q_\ell m_\ell \left(Y_{\ell, d}^2 - Y_{\ell, s}^2\right) {\cal{A}}_{HA,\ell} \right|^2 \,,
\end{equation}
where $\beta_H = \sqrt{1 - \frac{m_H^2}{4 m_{B^{(1)}}^2}}$.  

\begin{figure*}[t]
\begin{center}
\includegraphics[scale=0.8]{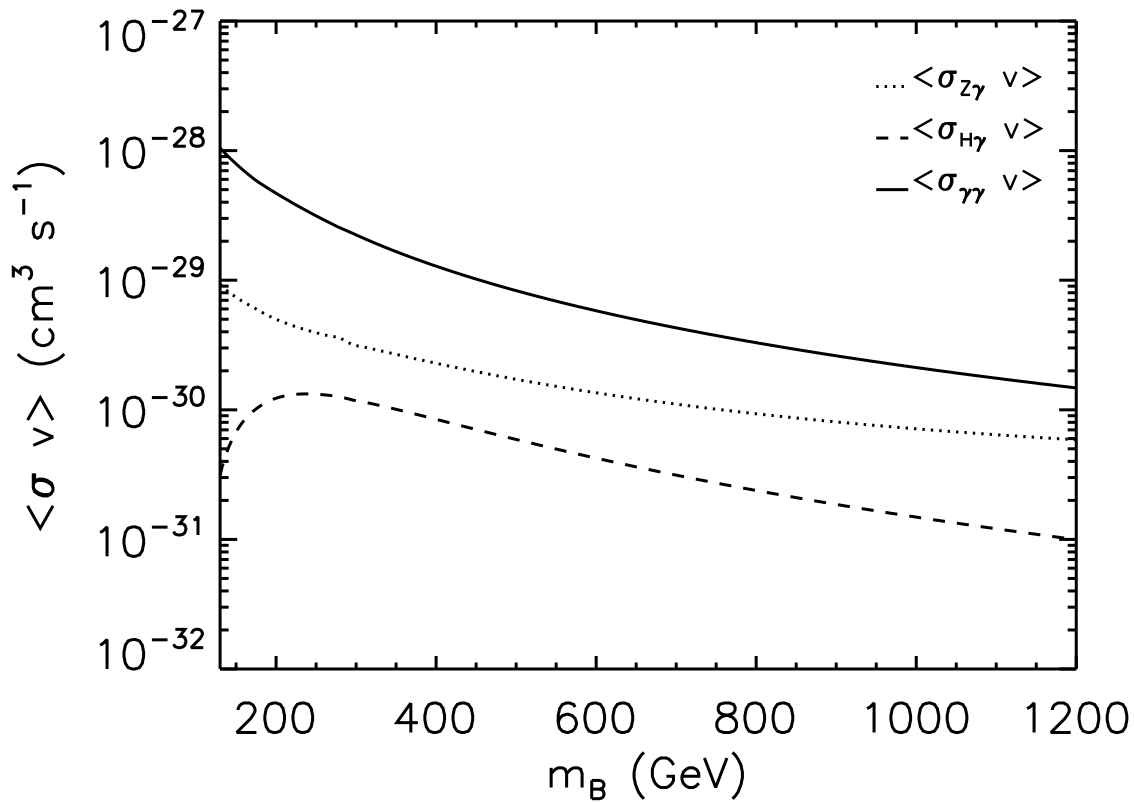}
\end{center}
\caption[]{The annihilation cross sections for the $\gamma\gamma$, $\gamma Z$ and $\gamma H$ final states.}
\label{fg:xn-vs-MB}
\end{figure*}

In Fig.~\ref{fg:xn-vs-MB}, we plot the cross sections for the $\gamma \gamma$, $\gamma Z$ and 
$\gamma H$ final states as a function of the WIMP mass $m_{B^{(1)}}$ where we have assumed a 
universal KK fermion mass of $m_\xi \simeq 1.1 \times m_{B^{(1)}}$.  From this figure, we see that 
the $\gamma Z$ cross section is roughly 10\% of the $\gamma \gamma$ cross section and is in 
agreement with the estimation of Ref.~\cite{Bergstrom:2004nr} while the $\gamma H$ cross section 
is even more suppressed.

\section{Gamma ray flux and energy spectrum}
\label{sec:flux}

The differential flux of photons predicted to be observed at an angle $\psi$ from the direction to the galactic center (GC) is 

\begin{equation}
\frac{d\Phi_{\gamma}}{d\Omega dE}(\psi,E)=\frac{r_{\odot}\rho_{\odot}^2}{4\pi m_{B^{(1)}}^2}\frac{dN_{\gamma}}{dE} \int_{\textrm{l.o.s.}}\frac{ds}{r_{\odot}}\,\left[\frac{\rho[r(s,\psi)]}{\rho_{\odot}}\right]^2\label{eq:diff_spectr}
\end{equation} 
with 
\begin{equation}
\frac{dN_{\gamma}}{dE}=\sum_f\langle\sigma v\rangle_f \frac{dN_{\gamma}^f}{dE},
\end{equation}
where the index $f$ denotes the annihilation channels with one or more photons in the final state, 
$\langle\sigma v\rangle_f$ is the corresponding cross-section and $dN_{\gamma}^f/dE$ is the 
(normalized) photon spectrum per annihilation. Furthermore, 
$\rho(\vec{x})$, $\rho_{\odot}=0.3$ GeV/cm$^3$ and $r_{\odot}=8.5$ kpc respectively 
denote the dark matter density at a generic location $\vec{x}$ \textit{with respect to the GC}, 
its value at the solar system location and the distance of the Sun from the GC. Finally, the 
coordinate $s$ runs along the line of sight, which in turn makes an angle $\psi$ with respect 
to the direction of the GC.

Equation (\ref{eq:diff_spectr}) allows one to separate the factors contributing to the predicted 
incoming gamma ray flux. In particular, while the $dN_{\gamma}/dE$ term is sensitive only to the 
particle physics under consideration, the remaining factors are sensitive to the modeling of the halo 
density profile $\rho(\vec{x})$.  For a given dark matter model,
these latter factors are the main source of uncertainty in the prediction of the detectability of a dark 
matter signal. To proceed further, we define with $J$ the dimensionless integral along the line-of-
sight appearing in Eq.~(\ref{eq:diff_spectr}) and with $\bar{J}$ its average value computed for a 
solid angle $\Delta\Omega$ centered on the GC
\begin{eqnarray}
J&\equiv&\int_{\textrm{l.o.s.}}
\frac{ds}{r_{\odot}}\,\left[\frac{\rho[r(s,\psi)]}{\rho_{\odot}}\right]^2,\nonumber\\
\bar{J}(\Delta\Omega)&=&\frac{1}{\Delta\Omega}\int_{\Delta\Omega}J(\psi)\,d\Omega.
\end{eqnarray}
The $\bar{J}$ factor thus defines the normalization of the gamma ray flux signal and allows one to 
quantify the impact of astrophysical uncertainties due to the lack of knowledge of the halo density 
profile. In what follows we will consider two models for $\rho$: the Navarro Frenk and White (NFW) 
profile and the ``adiabatically contracted'' profile. 

The Navarro Frenk and White (NFW) profile has been shown to fit reasonably well the results from 
recent high-resolution numerical simulations. This density profile is often used as a benchmark for 
indirect dark matter searches \cite{Navarro:1995iw},
\begin{equation}
\label{eq:NFW}
\rho_{\rm{NFW}}(r) = \frac{\rho_s}{\frac{r}{r_s}\left( 1 + \frac{r}{r_s}\right)^2} \,.
\end{equation}

Modifications of the above profile on very small scales 
have been observed in the most recent simulations.
While Ref.~\cite{Diemand:2005wv} shows that
the innermost regions of DM halos are better 
approximated with $r^{-1.2}$ cusps, Ref.~\cite{Navarro:2008kc} points out that
the analytic form that provides an optimal fit
to the simulated halos is the so-called ``Einasto profile" \cite{Graham:2005xx},
\begin{equation}
\rho(r)=\rho_0\,\exp\left[-\frac{2}{\alpha}\left(\left(\frac{r}{R}\right)^{\alpha}-1\right)\right],
\end{equation} 
which is shallower than NFW at very small radii\footnote{The values assumed for the parameters are in this case $\alpha=0.17$ and $R=20$ kpc.}.

The above results have been derived in the framework of simulations containing only dark matter particles, which interact gravitationally. 
Baryons however are expected to have a non-negligible effect on the small scale dynamics of galaxies. In particular, due to the dissipative nature of the baryonic fluid, when baryons lose energy and contract, this affects the gravitational potential experienced by the dark matter. In the ``adiabatic compression" scenario~\cite{Blumenthal:1985qy}, 
the baryons contraction is quasi-stationary and spherically symmetric.
Starting from an initial NFW profile, the final slope in the
innermost regions becomes $r^{-1.5}$~\cite{Edsjo:2004pf,Prada:2004pi,Gnedin:2004cx,Bertone:2005hw}.

Assuming $\Delta\Omega=10^{-5}$ sr. (corresponding to the angular resolution of the HESS and Fermi LAT $\gamma$-ray experiments), in Tab.~\ref{tab:DM_prof} we show the value of $\bar{J}$ obtained for two profiles: the benchmark NFW profile and the ``adiabatically contracted'' profile, with the same parameters as the profile labelled `A' in Ref.~\cite{Bertone:2005hw}\footnote{Note that our values of $J$ for the NFW and ``Adiabatic'' profiles are slightly different from the 
values in Ref. \cite{Bertone:2005hw},
due to the fact that the NFW profile was there approximated as a simple $r^{-1}$
power-law from the galactic center to the location of the Sun.}. Table \ref{tab:DM_prof} explicitly shows the extent to which the present uncertainty in the knowledge of the halo density profile turns into uncertainty in the predicted gamma ray flux from the GC. 
\begin{table}[t]
\begin{tabular}{c|c}
\hline
~~~ Model ~~~ & ~~~ $\bar{J}\left( 10^{-5}\right)$ ~~~ \\
\hline\hline
NFW& $1.5 \times 10^4$\\
Adiabatic & $4.7 \times 10^7$ \\
\hline
\end{tabular}
\caption{\label{tab:DM_prof} Value of $\bar{J}(10^{-5})$ for two dark matter density profiles.}
\end{table}

Recently, a large suite of cosmological simulations aimed to study the assembly of Milky Way like 
galaxies were carried out using the Adaptive Mesh Refinement (AMR) code {\tt RAMSES}, which 
includes treatment of dark matter, gas and stars \cite{ATM10} . The modeling included realistic 
recipes for star formation, supernova feedback (SNII and SNIa), stellar mass loss, gas cooling/
heating and metal enrichment. It was found that adopting different prescriptions for the star 
formation history results in a strong difference not only in galactic disk size and concentration, but 
also on DM contraction, therefore dramatically affecting the DM profile in the innermost regions of 
galactic halos. 

These simulations were shown to bracket the so-called Kennicutt-Schmidt relation (that 
parametrizes the star formation rate as a function of the gas density) from the THINGS survey \cite
{Bigiel} , and they are therefore representative of Milky Way-like galaxies in the local Universe. A 
comparison of the profiles obtained in these simulations with the standard DM-only case can be 
found e.g. in Ref. \cite{pato}, where it is also shown that the final profile is substantially different 
with respect to a na\"ive implementation of adiabatic contraction. We also stress that DM halos in 
numerical simulations deviate rather strongly from spherical symmetry, a circumstance that 
introduces additional uncertainties on the estimate of the integral along the line of sight 
$J$ \cite{pato}.

Other processes due to the gravitational interaction of baryons (such as the ``heating" of the DM 
fluid) can possibly have the opposite effect on the final DM distribution and soften the central 
cusps. Furthermore, the presence of a super-massive black hole (BH) at the GC inevitably affects 
the DM profile. 
The growth of the BH from an initial small seed would initially lead to a large DM ``spike'' 
\cite{Gondolo:1999ef}. While dynamical effects  and DM annihilations will subsequently tend to 
destroy the spike \cite{Merritt:2002vj,Ullio:2001fb,Bertone:2005hw}, 
in some cases significant overdensities can survive over a Hubble time. 

In what follows, the predicted gamma ray flux from the galactic center is obtained assuming the 
NFW halo profile. To obtain predictions for other profiles, it is sufficient to rescale the flux by the 
appropriate ratio of $\bar{J}$ (which is $3.3\cdot10^3$ for a na\"ive implementation of adiabatic 
contraction). 

Finally, we account for the finite resolution of the detector by convolving the unfiltered signal 
$S(E)$ with a gaussian kernel $G(E,E_0)$,
\begin{equation}
G(E,E_0)=\frac{1}{\sqrt{2\pi}E_0\sigma} \exp\left[-\frac{(E-E_0)^2}{2\sigma^2E_0^2}\right],
\end{equation} 
where $\sigma$ is related to the detector's relative energy resolution $\xi$ by $\sigma=\xi/2.3$. 
The signal $S_M(E_0)$ measured by the detector at energy $E_0$ is then simply given by
\begin{equation}
S_M(E_0)=\int dE\, G(E,E_0)\,S(E).
\end{equation}

\begin{figure*}[t]
\begin{center}
\includegraphics[scale=0.8]{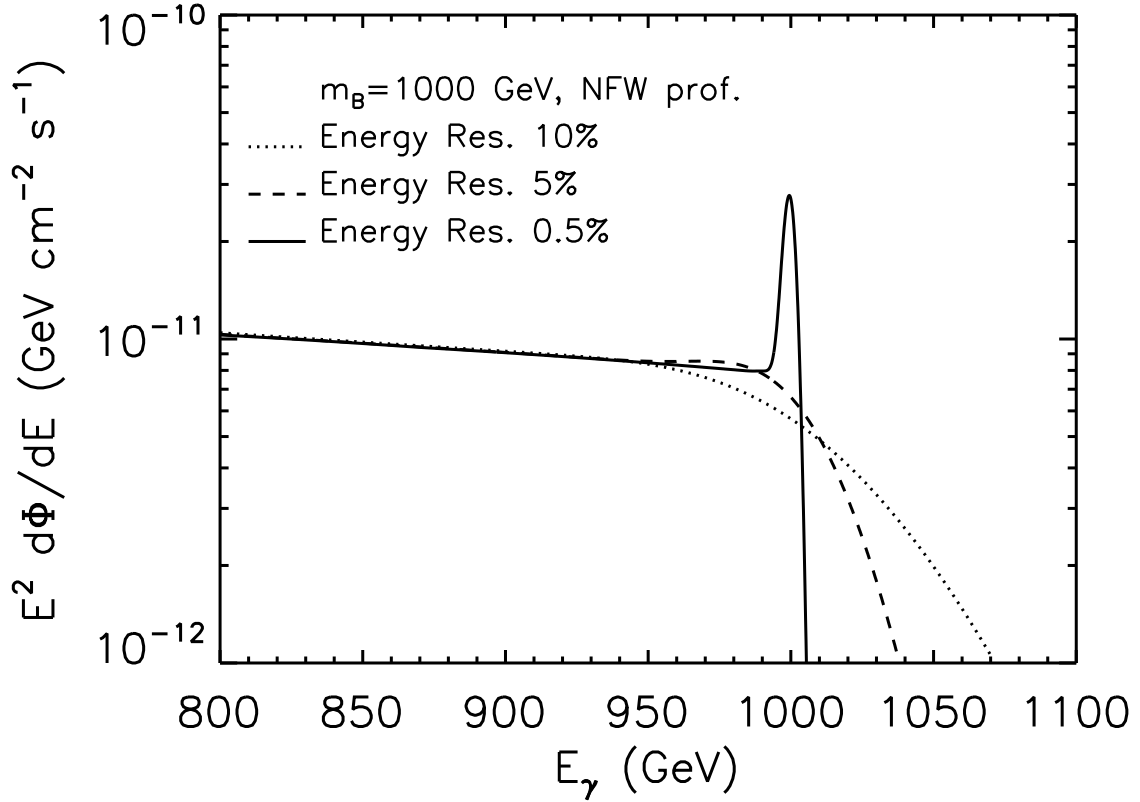}
\end{center}
\caption[]{The gamma ray flux as a function of the photon's energy for a WIMP of mass 1 TeV.  Shown are three different experimental energy resolutions.}
\label{fg:flux-MB1000}
\end{figure*} 

\begin{figure*}[t]
\begin{center}
\includegraphics[scale=0.8]{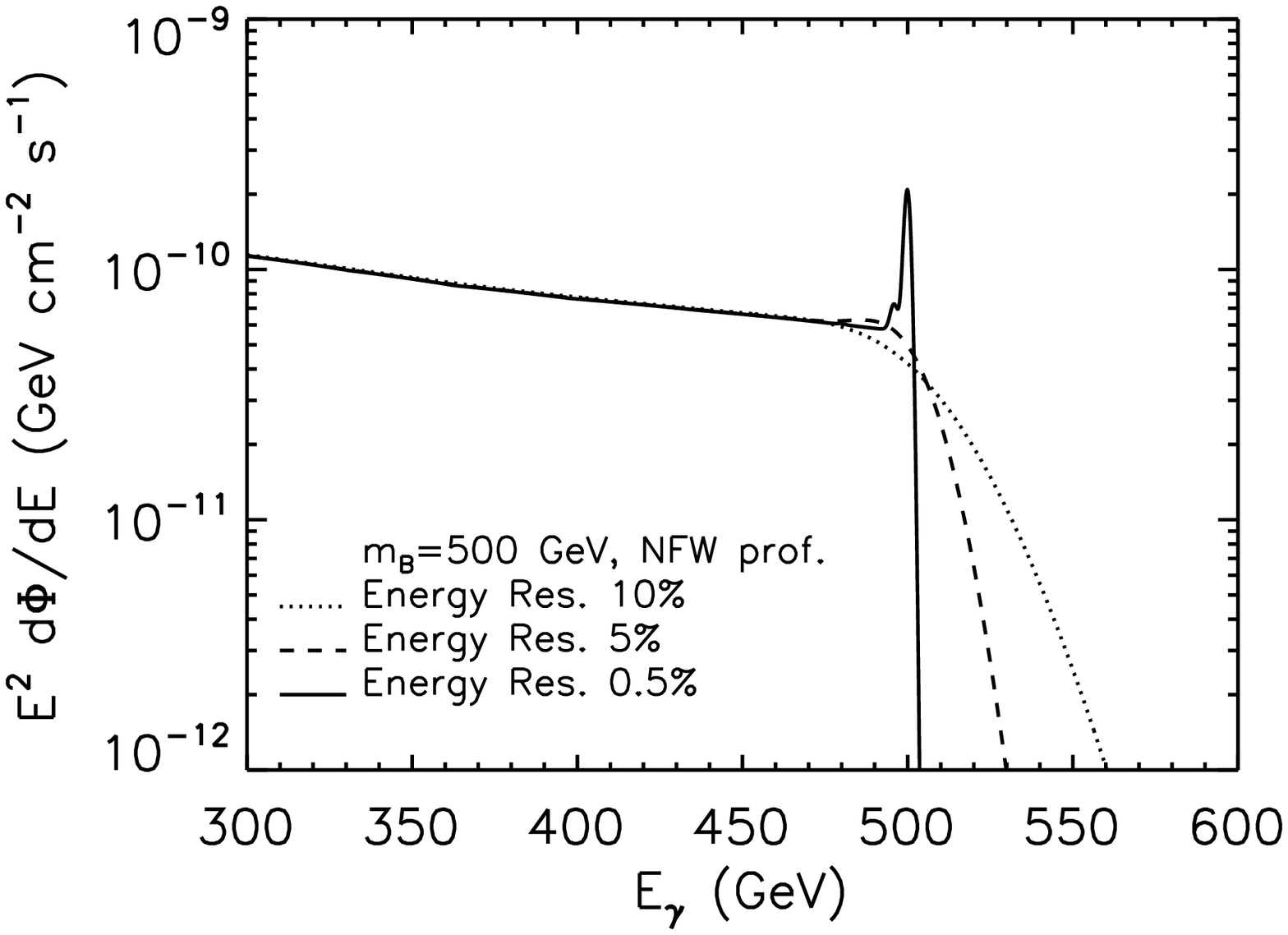}
\end{center}
\caption[]{The gamma ray flux as a function of the photon's energy for a WIMP of mass 500 GeV.  Shown are three different experimental energy resolutions.}
\label{fg:flux-MB500}
\end{figure*} 

\begin{figure*}[t]
\begin{center}
\includegraphics[scale=0.8]{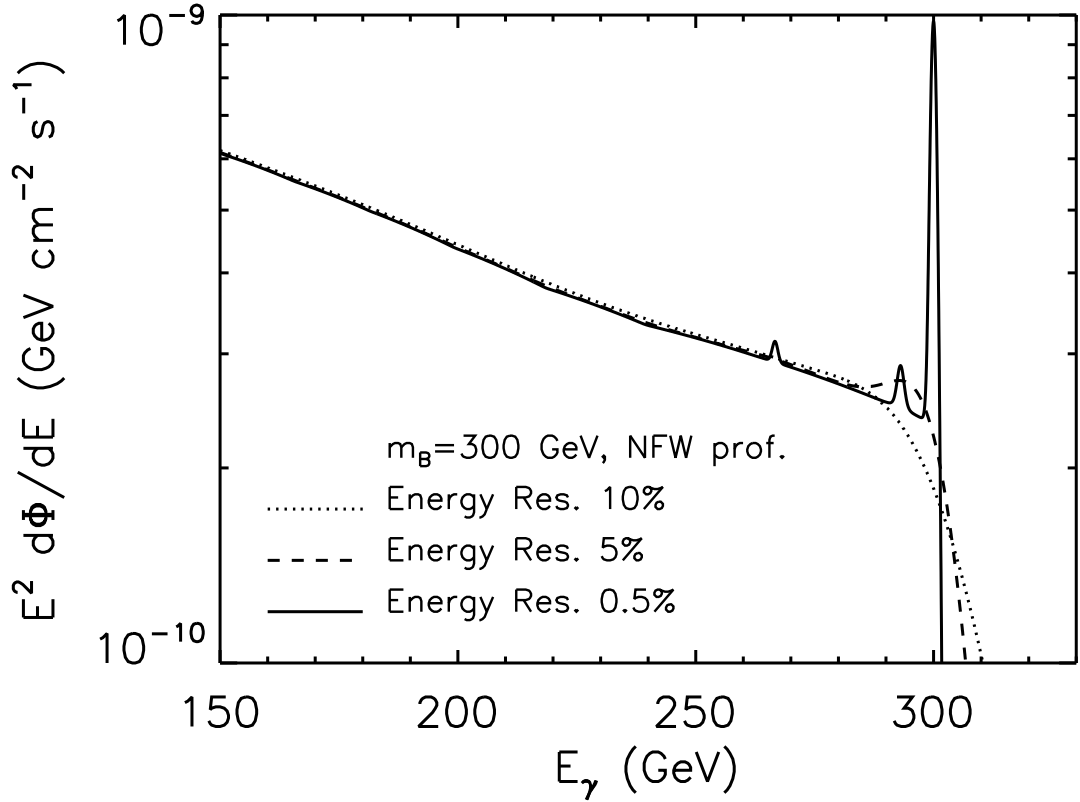}
\end{center}
\caption[]{The gamma ray flux as a function of the photon's energy for a WIMP of mass 300 GeV.  Shown are three different experimental energy resolutions.}
\label{fg:flux-MB300}
\end{figure*} 

Combining the expectations for $\bar{J}$ with the particle physics rates and the detector response,
we arrive at predictions for the flux and spectrum of gamma rays.
In Figs.~\ref{fg:flux-MB1000}, \ref{fg:flux-MB500} and \ref{fg:flux-MB300}, we show predictions
for the flux for an NFW halo profile and three choices of LKP mass.  Although the thermal
relic density favors LKPs in the range of 500 - 900 GeV \cite{Servant:2002aq,Kong:2005hn},
we consider a wider range of LKP masses motivated by the possibility that
the LKP is not a canonical thermal relic, which could allow for a wider
range of masses consistent with the observed quantity of dark matter in the Universe.
For example, if the modulus which controls the
size of the extra dimension is cosmologically active, it could cause large
deviations in the thermodynamics of the Universe from its extrapolated history \cite{Kolb:2003mm}.
Alternately, early production of KK gravitons which eventually decay into LKPs can serve
as a nonthermal production mechanism for KK dark matter in an otherwise standard
cosmological history \cite{Shah:2006gs}.

Figures~\ref{fg:flux-MB1000} -- \ref{fg:flux-MB300} contain curves for three different choices
of energy resolutions, ranging from the $\sim 10\%$ resolution typical of current experiments, to
an aggressive $0.5\%$ resolution which might be possible in future experiments.  We find
that at $10\%$ energy resolution, lines in the 5d UED model are very difficult to distinguish from
the continuum.  At a $5\%$ energy resolution, broad lines may appear for LKP masses 
around 300 GeV, slightly above the lower bound from colliders.  At $0.5\%$, well separated lines
for $\gamma \gamma$, $\gamma Z$, and $\gamma H$ are visible for light LKPs, and
some structure related to the $\gamma H$ line is visible for an LKP mass of around 500 GeV.

In principle, we should compare our predicted flux with gamma ray observations, since data are 
available from a variety of gamma-ray telescopes, such as the Fermi LAT and Air Cherenkov 
Telescopes like HESS and MAGIC. The comparison is however made complicated by the 
aforementioned uncertainties on the normalization of the predicted flux on one side, and on the 
other by the difficult interpretation of astrophysical data. 

It is for instance unclear at this stage whether known astrophysical sources can account for the 
energy spectrum obtained by Fermi. A recent analysis of a $1^\circ \times 1^\circ$ found that the 
diffuse gamma-ray background and discrete sources, as estimated with standard astrophysical 
tools, can account for practically all the observed flux, but there is a small excess at energies 
$\cal O$(1) GeV, which may or may not be explained with an improved model of the Galactic 
diffuse emission and with a better understanding of the systematic errors \cite{fermiGC}. At the 
same time, there is an ongoing debate on the possible origin of the point source observed by 
HESS right at the Galactic center \cite{hessGC}. 

In principle, once these issues are settled, and the astrophysical backgrounds are well-known, 
one can search for a DM signal even in the case where it provides a subdominant contribution to 
the total flux. 
More subtle features such as subdominant gamma-ray lines could therefore in principle emerge in 
the process of data analysis, in which case one could try to perform the program describe above, 
and search for additional lines that might shed light on the structure of the dark sector.

\section{conclusions}
\label{sec:conclusions}

Extra dimensions are a fascinating possibility in the spectrum of physics beyond the Standard
Model, one which leads to a unique mechanism to provide a stable WIMP to play the role of dark
matter.  We have examined the gamma ray line signals for UED dark matter WIMPs, extending
the results for $\gamma \gamma$ in the literature \cite{Bergstrom:2004nr} to also include lines
from $\gamma Z$ and $\gamma H$ final states.
Our conclusions are that such lines are very challenging for the current generation of gamma
ray observatories to resolve, but may be visible at the next generation of such experiments.  As
such experiments are designed, we hope our work will be of some value in evaluating their
exciting physics capabilities.

\begin{acknowledgments}
\end{acknowledgments}
T Tait is grateful to the SLAC theory group for their extraordinary
generosity during his many visits and is partially supported by 
NSF grant PHY-0970171.
Research at Argonne National Laboratory is
supported in part by the Department of Energy under contract
DE-AC02-06CH11357.  GS is also supported in part by the Department of Energy under contract DE-FG02-91ER40684.  AV is supported by DOE at Fermilab. AV thanks Fermilab Center for Particle Astrophysics for hospitality during the final stages of this work.


\begin{thebibliography}{99}

\bibitem{Bertone:2004pz}
  G.~Bertone, D.~Hooper and J.~Silk,
  Phys.\ Rept.\  {\bf 405}, 279 (2005)
  [arXiv:hep-ph/0404175].

\bibitem{Appelquist:2000nn}
  T.~Appelquist, H.~C.~Cheng and B.~A.~Dobrescu,
  Phys.\ Rev.\  D {\bf 64}, 035002 (2001)
  [arXiv:hep-ph/0012100].
  
\bibitem{Servant:2002aq}
  G.~Servant and T.~M.~P.~Tait,
  Nucl.\ Phys.\  B {\bf 650}, 391 (2003)
  [arXiv:hep-ph/0206071].
  
\bibitem{Kong:2005hn}
  K.~Kong and K.~T.~Matchev,
  JHEP {\bf 0601}, 038 (2006)
  [arXiv:hep-ph/0509119];
  F.~Burnell and G.~D.~Kribs,
  Phys.\ Rev.\  D {\bf 73}, 015001 (2006)
  [arXiv:hep-ph/0509118];
  M.~Kakizaki, S.~Matsumoto, Y.~Sato and M.~Senami,
  Nucl.\ Phys.\  B {\bf 735}, 84 (2006)
  [arXiv:hep-ph/0508283];
  M.~Kakizaki, S.~Matsumoto and M.~Senami,
  Phys.\ Rev.\  D {\bf 74}, 023504 (2006)
  [arXiv:hep-ph/0605280];
  S.~Matsumoto, J.~Sato, M.~Senami and M.~Yamanaka,
  Phys.\ Rev.\  D {\bf 76}, 043528 (2007)
  [arXiv:0705.0934 [hep-ph]].

\bibitem{Appelquist:2002wb}
  T.~Appelquist and H.~U.~Yee,
  Phys.\ Rev.\  D {\bf 67}, 055002 (2003)
  [arXiv:hep-ph/0211023];
  T.~Flacke, D.~Hooper and J.~March-Russell,
  Phys.\ Rev.\  D {\bf 73}, 095002 (2006)
  [Erratum-ibid.\  D {\bf 74}, 019902 (2006)]
  [arXiv:hep-ph/0509352].
  I.~Gogoladze and C.~Macesanu,
  Phys.\ Rev.\  D {\bf 74}, 093012 (2006)
  [arXiv:hep-ph/0605207].
    
\bibitem{Cheng:2002ab}
  H.~C.~Cheng, K.~T.~Matchev and M.~Schmaltz,
  Phys.\ Rev.\  D {\bf 66}, 056006 (2002)
  [arXiv:hep-ph/0205314].

\bibitem{Datta:2005zs}
  A.~Datta, K.~Kong and K.~T.~Matchev,
  Phys.\ Rev.\  D {\bf 72}, 096006 (2005)
  [Erratum-ibid.\  D {\bf 72}, 119901 (2005)]
  [arXiv:hep-ph/0509246];
  J.~A.~R.~Cembranos, J.~L.~Feng and L.~E.~Strigari,
  Phys.\ Rev.\  D {\bf 75}, 036004 (2007)
  [arXiv:hep-ph/0612157];
  S.~Matsumoto, J.~Sato, M.~Senami and M.~Yamanaka,
  Phys.\ Rev.\  D {\bf 80}, 056006 (2009)
  [arXiv:0903.3255 [hep-ph]];
  G.~Bhattacharyya, A.~Datta, S.~K.~Majee and A.~Raychaudhuri,
  Nucl.\ Phys.\  B {\bf 821}, 48 (2009)
  [arXiv:0904.0937 [hep-ph]];
  P.~Bandyopadhyay, B.~Bhattacherjee and A.~Datta,
  JHEP {\bf 1003}, 048 (2010)
  [arXiv:0909.3108 [hep-ph]];
  B.~Bhattacherjee and K.~Ghosh,
  arXiv:1006.3043 [hep-ph].
  
\bibitem{Cheng:2002ej}
  H.~C.~Cheng, J.~L.~Feng and K.~T.~Matchev,
  Phys.\ Rev.\ Lett.\  {\bf 89}, 211301 (2002)
  [arXiv:hep-ph/0207125].
  
\bibitem{Servant:2002hb}
  G.~Servant and T.~M.~P.~Tait,
  New J.\ Phys.\  {\bf 4}, 99 (2002)
  [arXiv:hep-ph/0209262].
  
\bibitem{Baltz:2004ie}
  E.~A.~Baltz and D.~Hooper,
  JCAP {\bf 0507}, 001 (2005)
  [arXiv:hep-ph/0411053];
  D.~Hooper and J.~Silk,
  Phys.\ Rev.\  D {\bf 71}, 083503 (2005)
  [arXiv:hep-ph/0409104];
  D.~Hooper and G.~D.~Kribs,
  Phys.\ Rev.\  D {\bf 70}, 115004 (2004)
  [arXiv:hep-ph/0406026].
  
\bibitem{Hooper:2002gs}
  D.~Hooper and G.~D.~Kribs,
  Phys.\ Rev.\  D {\bf 67}, 055003 (2003)
  [arXiv:hep-ph/0208261];
  A.~E.~Erkoca, M.~H.~Reno and I.~Sarcevic,
  arXiv:1009.2068 [hep-ph].

\bibitem{Barrau:2005au}
  A.~Barrau, P.~Salati, G.~Servant, F.~Donato, J.~Grain, D.~Maurin and R.~Taillet,
  Phys.\ Rev.\  D {\bf 72}, 063507 (2005)
  [arXiv:astro-ph/0506389];
  T.~Bringmann,
  JCAP {\bf 0508}, 006 (2005)
  [arXiv:astro-ph/0506219].
  
\bibitem{Bertone:2002ms}
  G.~Bertone, G.~Servant and G.~Sigl,
  Phys.\ Rev.\  D {\bf 68}, 044008 (2003)
  [arXiv:hep-ph/0211342].
  L.~Bergstrom, T.~Bringmann, M.~Eriksson and M.~Gustafsson,
  Phys.\ Rev.\ Lett.\  {\bf 94}, 131301 (2005)
  [arXiv:astro-ph/0410359].
  
\bibitem{Bergstrom:2004nr}
  L.~Bergstrom, T.~Bringmann, M.~Eriksson and M.~Gustafsson,
  JCAP {\bf 0504}, 004 (2005)
  [arXiv:hep-ph/0412001].
  
\bibitem{Mack:2008wu}
  G.~D.~Mack, T.~D.~Jacques, J.~F.~Beacom, N.~F.~Bell and H.~Yuksel,
  Phys.\ Rev.\  D {\bf 78}, 063542 (2008)
  [arXiv:0803.0157 [astro-ph]].
  
\bibitem{Abdo:2010nc}
  A.~A.~Abdo {\it et al.},
  Phys.\ Rev.\ Lett.\  {\bf 104}, 091302 (2010)
  [arXiv:1001.4836 [astro-ph.HE]].
  
\bibitem{Bertone:2009cb}
  G.~Bertone, C.~B.~Jackson, G.~Shaughnessy, T.~M.~P.~Tait and A.~Vallinotto,
  Phys.\ Rev.\  D {\bf 80}, 023512 (2009)
  [arXiv:0904.1442 [astro-ph.HE]].
  
 \bibitem{Bergstrom:1997fh}
  L.~Bergstrom and P.~Ullio,
  Nucl.\ Phys.\  B {\bf 504}, 27 (1997)
  [arXiv:hep-ph/9706232];
  Z.~Bern, P.~Gondolo and M.~Perelstein,
  Phys.\ Lett.\  B {\bf 411}, 86 (1997)
  [arXiv:hep-ph/9706538];
  P.~Ullio and L.~Bergstrom,
  Phys.\ Rev.\  D {\bf 57}, 1962 (1998)
  [arXiv:hep-ph/9707333];
 L.~Bergstrom, P.~Ullio and J.~H.~Buckley,
 Astropart.\ Phys.\  {\bf 9}, 137 (1998)
 [arXiv:astro-ph/9712318];
  F.~Boudjema, A.~Semenov and D.~Temes,
  Phys.\ Rev.\  D {\bf 72}, 055024 (2005)
  [arXiv:hep-ph/0507127];

\bibitem{Dudas:2009uq}
  E.~Dudas, Y.~Mambrini, S.~Pokorski and A.~Romagnoni,
  arXiv:0904.1745 [hep-ph];
  Y.~Mambrini,
  JCAP {\bf 0912}, 005 (2009)
  [arXiv:0907.2918 [hep-ph]];
  C.~B.~Jackson, G.~Servant, G.~Shaughnessy, T.~M.~P.~Tait and M.~Taoso,
  JCAP {\bf 1004}, 004 (2010)
  [arXiv:0912.0004 [hep-ph]].
  C.~Arina, T.~Hambye, A.~Ibarra and C.~Weniger,
  JCAP {\bf 1003}, 024 (2010)
  [arXiv:0912.4496 [hep-ph]].

\bibitem{Perelstein:2006bq}
  M.~Perelstein and A.~Spray,
  Phys.\ Rev.\  D {\bf 75} (2007) 083519
  [arXiv:hep-ph/0610357];
  M.~Gustafsson, E.~Lundstrom, L.~Bergstrom and J.~Edsjo,
  Phys.\ Rev.\ Lett.\  {\bf 99} (2007) 041301
  [arXiv:astro-ph/0703512].
  
\bibitem{Goodman:2010qn}
  J.~Goodman, M.~Ibe, A.~Rajaraman, W.~Shepherd, T.~M.~P.~Tait and H.~B.~P.~Yu,
  arXiv:1009.0008 [hep-ph].
  
\bibitem{Georgi:2000ks}
  H.~Georgi, A.~K.~Grant and G.~Hailu,
  Phys.\ Lett.\  B {\bf 506}, 207 (2001)
  [arXiv:hep-ph/0012379].
  
\bibitem{Cheng:2002iz}
  H.~C.~Cheng, K.~T.~Matchev and M.~Schmaltz,
  Phys.\ Rev.\  D {\bf 66}, 036005 (2002)
  [arXiv:hep-ph/0204342].
  
\bibitem{Carena:2002me}
  M.~S.~Carena, T.~M.~P.~Tait and C.~E.~M.~Wagner,
  Acta Phys.\ Polon.\  B {\bf 33}, 2355 (2002)
  [arXiv:hep-ph/0207056].
  
\bibitem{Flacke:2008ne}
  T.~Flacke, A.~Menon and D.~J.~Phalen,
  Phys.\ Rev.\  D {\bf 79}, 056009 (2009)
  [arXiv:0811.1598 [hep-ph]];
  C.~Csaki, J.~Heinonen, J.~Hubisz, S.~C.~Park and J.~Shu,
  arXiv:1007.0025 [hep-ph].
  
\bibitem{Passarino:1978jh}
  G.~Passarino and M.~J.~G.~Veltman,
  Nucl.\ Phys.\  B {\bf 160}, 151 (1979).
  
\bibitem{Stuart:1987tt}
  R.~G.~Stuart,
  Comput.\ Phys.\ Commun.\  {\bf 48}, 367 (1988).
  
\bibitem{Navarro:1995iw}
 J.~F.~Navarro, C.~S.~Frenk and S.~D.~M.~White,
 Astrophys.\ J.\  {\bf 462}, 563 (1996)
 [arXiv:astro-ph/9508025].
 
\bibitem{Diemand:2005wv}
  J.~Diemand, M.~Zemp, B.~Moore, J.~Stadel and M.~Carollo,
  Mon.\ Not.\ Roy.\ Astron.\ Soc.\  {\bf 364}, 665 (2005)
  [arXiv:astro-ph/0504215].
  
\bibitem{Navarro:2008kc}
  J.~F.~Navarro {\it et al.},
  arXiv:0810.1522 [astro-ph].

\bibitem{Graham:2005xx}
 A.~W.~Graham, D.~Merritt, B.~Moore, J.~Diemand and B.~Terzic,
 Astron.\ J.\  {\bf 132}, 2685 (2006)
 [arXiv:astro-ph/0509417].

\bibitem{Blumenthal:1985qy}
 G.~R.~Blumenthal, S.~M.~Faber, R.~Flores and J.~R.~Primack,
 Astrophys.\ J.\  {\bf 301}, 27 (1986).

\bibitem{Edsjo:2004pf}
 J.~Edsjo, M.~Schelke and P.~Ullio,
 JCAP {\bf 0409}, 004 (2004)
 [arXiv:astro-ph/0405414].

\bibitem{Prada:2004pi}
 F.~Prada, A.~Klypin, J.~Flix, M.~Martinez and E.~Simonneau,
 Phys.\ Rev.\ Lett.\  {\bf 93}, 241301 (2004)
 [arXiv:astro-ph/0401512].

\bibitem{Gnedin:2004cx}
 O.~Y.~Gnedin, A.~V.~Kravtsov, A.~A.~Klypin and D.~Nagai,
 Astrophys.\ J.\  {\bf 616}, 16 (2004)
 [arXiv:astro-ph/0406247].
 
\bibitem{Bertone:2005hw}
 G.~Bertone and D.~Merritt,
 Phys.\ Rev.\  D {\bf 72} (2005) 103502
 [arXiv:astro-ph/0501555].

  \bibitem{ATM10}
  O.~Agertz, R.~Teyssier and B.~Moore,
  arXiv:1004.0005 [Unknown].

\bibitem{Bigiel}
  F.~Bigiel, A.~Leroy, F.~Walter, E.~Brinks, W.~J.~G.~de Blok, B.~Madore and M.~D.~Thornley,
  Astron.\ J.\  {\bf 136} (2008) 2846
  [arXiv:0810.2541 [astro-ph]].

\bibitem{pato}
  M.~Pato, O.~Agertz, G.~Bertone, B.~Moore and R.~Teyssier,
  Phys.\ Rev.\  D {\bf 82} (2010) 023531
  [arXiv:1006.1322 [astro-ph.HE]].

\bibitem{Gondolo:1999ef}
 P.~Gondolo and J.~Silk,
 Phys.\ Rev.\ Lett.\  {\bf 83}, 1719 (1999)
 [arXiv:astro-ph/9906391].

\bibitem{Merritt:2002vj}
 D.~Merritt, M.~Milosavljevic, L.~Verde and R.~Jimenez,
 Phys.\ Rev.\ Lett.\  {\bf 88}, 191301 (2002)
 [arXiv:astro-ph/0201376].

\bibitem{Ullio:2001fb}
 P.~Ullio, H.~Zhao and M.~Kamionkowski,
 Phys.\ Rev.\  D {\bf 64}, 043504 (2001)
 [arXiv:astro-ph/0101481].

\bibitem{Kolb:2003mm}
  E.~W.~Kolb, G.~Servant and T.~M.~P.~Tait,
  JCAP {\bf 0307}, 008 (2003)
  [arXiv:hep-ph/0306159];
  K.~C.~Chan and M.~C.~Chu,
  Phys.\ Rev.\  D {\bf 77}, 063525 (2008)
  [arXiv:0708.0122 [astro-ph]].
  
\bibitem{Shah:2006gs}
  N.~R.~Shah and C.~E.~M.~Wagner,
  Phys.\ Rev.\  D {\bf 74}, 104008 (2006)
  [arXiv:hep-ph/0608140].


\bibitem{fermiGC}
  V.~Vitale, A.~Morselli and f.~t.~F.~Collaboration,
  arXiv:0912.3828 [astro-ph.HE].
  
\bibitem{hessGC}
  H.~E.~S.~Aharonian,
  Astron. \& Astrophys. {\bf 503}, (2009) 817 
  [arXiv:0906.1247 [astro-ph.GA]].
  
  
\end{thebibliography}
\end{document}